\documentstyle[mprocl]{article}

\input{epsf}
\def\Journal#1#2#3#4{{#1} {\bf #2}, #3 (#4)}

\def\NPB{{\em Nucl. Phys.} B}
\def\PLB{{\em Phys. Lett.}  B}

\def\PRD{{\em Phys. Rev.} D}

\def\be{\begin{equation}}
\def\ee{\end{equation}}
\def\bea{\begin{eqnarray}}
\def\eea{\end{eqnarray}}

\def\CQG{\em Class.\ Quantum Grav.}

\begin{document}

\title{COMPLETE CLASSIFICATION OF 1+1 GRAVITY SOLUTIONS}

\author{ T. KL\"OSCH }

\address{Institut f\"ur Theoretische Physik, TU Wien,
Wiedner Hauptstr. 8--10,\\ A-1040 Vienna, Austria}

\author{ T. STROBL }

\address{Institut f\"ur Theoretische Physik, RWTH-Aachen,
Sommerfeldstr. 26--28,\\ D52056 Aachen, Germany}

\maketitle\abstracts{
A classification of the maximally extended solutions for 1+1
gravity models (comprising e.g.\ generalized dilaton gravity as well as
models with non-trivial torsion) is presented.
No restrictions are placed on the topology of the arising solutions,
and indeed it is found that for generic models solutions on
non-compact surfaces of arbitrary genus with an arbitrary non-zero number of
holes can be obtained. The moduli space of classical solutions (solutions of
the field equations with fixed topology modulo gauge transformations)
is parametrized explicitly.}
  
\section{Introduction}

A feature shared by practically all matterless 2D gravity models
is the presence of a Killing symmetry.
The models of this type comprise all 2D dilaton gravity
theories,\cite{Dil}
\begin{equation}
  L[g,\Phi] = \int_{\cal M} {\rm d}^2 x \sqrt{|\det g|}
                \big[U(\Phi) R + V(\Phi)
                   + W(\Phi) \partial_\mu \Phi \partial^\mu \Phi
                \big] \, ,
  \label{model}
\end{equation}
thus including e.g.\ spherically reduced gravity or $R^2$-gravity,
but also generalizations with non-trivial torsion, not contained in
(\ref{model}).\cite{CQG}
The classification of solutions splits naturally into two parts:
local considerations, i.e.\ solving the field equations for the metric and
other fields on some local patch, and global ones, like finding the maximal
extensions, determining their topology and causal structure.

\section{Local Issues}\label{sec:LocIs}

Due to the Killing symmetry
the metric may locally always be brought into Schwarz\-schild (SS)
or equally well into Eddington-Finkelstein (EF) form \cite{CQG,KuSchw}
\begin{equation}
  g^{S\!S}=h(r){\rm d}t^2-\frac{1}{h(r)}{\rm d}r^2 \,, \qquad
  g^{E\!F}=2{\rm d}r{\rm d}v+h(r){\rm d}v^2 \,,
 \label{metric}
\end{equation}
the function $h(r)$ being the same in both cases. In contrast to the
SS-form, the EF-form covers smoothly also the Killing horizons
(zeros of $h(r)$), allowing for a simple extension algorithm (see below).
Other fields of physical relevance like the dilaton $\Phi$ also
depend on $r$ only. Thus, locally the classification amounts to specifying
the possible functions $h(r)$ and $\Phi(r)$. 
Solving the field equations for these functions is relatively straightforward
and leads, for a fixed Lagrangian (\ref{model}), to a one-parameter family
$h=h_M(r)$, where $M$ can often be given a physical interpretation as
black-hole mass.

\section{Global Issues}

Unless $h(r)$ does not have zeros, the EF-patches (\ref{metric}) are
incomplete and thus have to be extended. 
Usually, for a solution to be maximally extended one expects that
geodesics should be complete or, if not, the curvature or some
other physical field like the dilaton $\Phi$ should blow up along them.
These requirements are not only highly reasonable physically,
solutions of this type also allow for a concise classification.
Unfortunately, there are other types of inextendibility, 
namely if the extension candidate\\
\noindent --- would not be Hausdorff (Taub-NUT spaces),\\
\noindent --- would not be smooth (conical singularity).\\
Especially solutions of the latter type abound: Just take {\em any\/} manifold,
cut out a point and take a covering manifold thereof. Clearly, the removed
point cannot be inserted any longer, since it would lead to a conical
singularity (branch point). Several popular kink metrics given in
the literature, but also recently discovered continuous families
of kinks are exactly of this type.\cite{Kink}
In what follows, however, we will disregard such solutions.

The extensions are best dealt with in two steps, constructing
the simply connected {\em universal coverings\/} (u.\ cov.) first.
To this end one exploits the symmetry of the metric under inversion
of the Killing parameter, $t\leftrightarrow -t$ in SS-coordinates.
In EF-coordinates this transformation reads
\begin{equation}
  r \mapsto r \,, \qquad v \mapsto -v-2\int\frac{{\rm d}r}{h(r)} \,,
 \label{flip}
\end{equation}
valid between two successive zeros of $h(r)$.
Using (\ref{flip}) as transition map for overlapping EF-patches and adding
special maps for the neighbourhood of bifurcation points, one obtains an
atlas for the u.\ cov.\cite{CQG}
This construction turns out to be unique for a given $h(r)$.
The corresponding multiply connected solutions can all
be obtained in a second step, by factoring this u.\ cov.\ by
adequately acting subgroups $\cal H$ of the symmetry group $\cal G$.
Moreover, the inequivalent factor spaces are in one-to-one correspondence
with the conjugacy classes of subgroups ${\cal H}\le{\cal G}$.

\begin{figure}[t]
\epsfxsize \textwidth \epsfbox{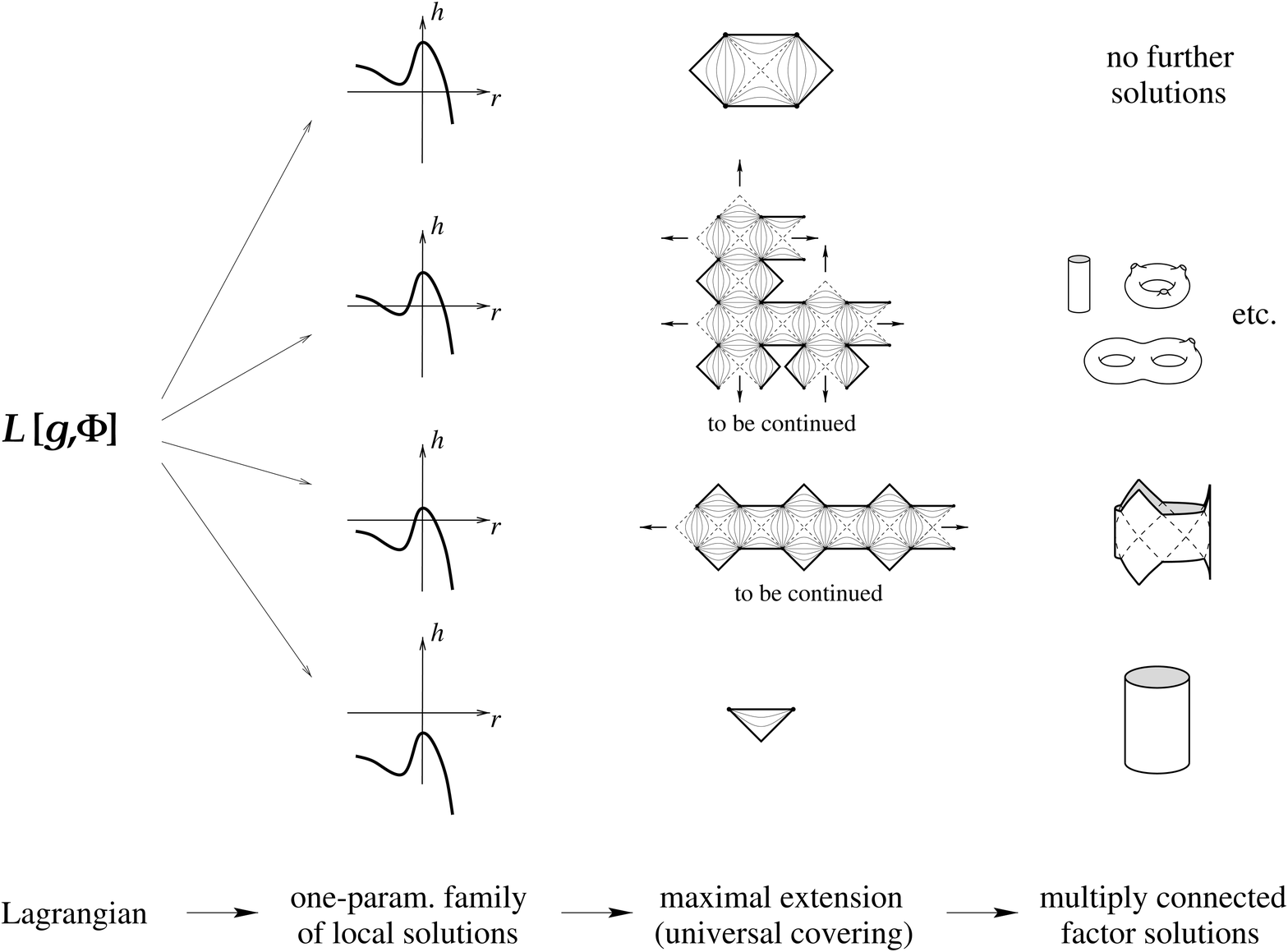}
\caption{Classification. A typical example.
\label{fig}}
\end{figure}

Applying the above program to a given metric (\ref{metric}), one finds the
global properties of the arising solutions to depend only on the number
and degree of the zeros of $h(r)$. Let us summarize the results for
space- and time-orientable solutions with merely
simple zeros of $h(r)$:\hspace{.1cm}\cite{CQG}
\begin{itemize}
\item there is a unique simply connected solution (universal covering),
 which is topologically a disc,
\item for no resp.\ two simple zeros of $h(r)$ there occur additionally
 cylinders,
\item for $\ge 3$ simple zeros there occur solutions of arbitrary non-compact
 topology,
\item the number of additional continuous parameters (besides $M$)
 equals the rank of $\pi_1(\mbox{spacetime manifold $\cal M$})$.
\end{itemize}
Note, however, that even within {\em one\/} model, i.e.\ for a fixed
Lagrangian, the number of zeros of $h(r)\equiv h_M(r)$ may change with the
parameter $M$ \nopagebreak (see e.g.\ Fig.\ \ref{fig}). Especially, for a
``generic'' choice of $U,V,W$ in the action (\ref{model}), the
occurrence of three or more zeros is the rule, resulting in solutions of
all non-compact topologies. For a fixed topology of $\cal M$
the moduli space, if non-empty, has dimension $\pi_1({\cal M}) + 1$.

\section*{Acknowledgments}

This work has been supported by the Austrian Fonds zur F\"orderung
der wissenschaftlichen Forschung (FWF), project P10221-PHY.


\section*{References}
\begin{thebibliography}{99}
\bibitem{Dil} T. Banks and M. O'Loughlin,
   \Journal{\NPB}{362}{649}{1991};\\
 S.D. Odintsov and I.L. Shapiro,
   \Journal{\PLB}{263}{183}{1991};\\
 D. Louis-Martinez,
 J. Gegenberg, G. Kunstatter,
   \Journal{\PLB}{321}{193}{1994}.
\bibitem{CQG} T. Kl\"osch and T. Strobl,
   \Journal{\CQG}{13}{965}{1996};\\
   \Journal{\em ibid.}{13}{2395}{1996};
   \Journal{\em ibid.}{14}{1689}{1997}.
\bibitem{KuSchw} W. Kummer and D.J. Schwarz,
   \Journal{\PRD}{45}{3628}{1992}.
\bibitem{Kink} T. Kl\"osch and T. Strobl, {\em A Global View of Kinks
   in 1+1 Gravity}, to appear in \PRD, gr-qc/9707053.

\end{thebibliography}
\end{document}